------------------------------------------------------------------------------

Appended: paper in LaTeX format (one file) and hep93.sty (one file)
          (see 'cut here')

--- Cut here ------------------------------------------------------------------

\documentstyle[epsf,hep93]{article}
\newcommand{\beq}{\begin{equation}}
\newcommand{\eeq}{\end{equation}}
\newcommand{\bea}{\begin{eqnarray}}
\newcommand{\eea}{\end{eqnarray}}

\newcommand{\half}{\textstyle {1\over2} \displaystyle}    % One half
\newcommand{\quarter}{\textstyle {1\over4} \displaystyle} % One quarter
\newcommand{\Dslash}{{\hbox{D}\kern-0.6em\raise0.15ex\hbox{/}}} % D slash

\renewcommand{\d}{\delta}

\newcommand{\cR}{{\cal R}}

\renewcommand{\thefootnote}{\fnsymbol{footnote}}
\begin{document}
\title{ SCALAR FIELDS COUPLED TO FOUR-DIMENSIONAL LATTICE GRAVITY
        \thanks{Talk presented at the Europhysics Conference on High Energy
        Physics, Marseille, France, July 22-28, 1993.}}
\author{HERBERT W. HAMBER \\
        \it Department of Physics, University of California at Irvine \\
        \it Irvine, Ca 92717, USA
        \
        }
\abstract{\rightskip=1.5pc
          \leftskip=1.5pc
I discuss some results we have obtained recently in a lattice model
for quantized gravity coupled to scalar matter in four dimensions.
We have looked at how the continuous phase transition
separating the smooth from the rough phase of gravity
is influenced by the presence of the scalar field.
We find that close to the critical point, where the average curvature
approaches zero, the effects of the scalar field are small and
the coupling of matter to gravity seems to be weak.
The nature of the phase diagram and the values for the critical
exponents would suggest that gravitational interactions increase with
distance. }

\maketitle
\vskip 5mm
{\bf \noindent Introduction}
\vskip 2.5mm

Of course any serious attempt at understanding the ground state properties
of quantized gravity has to include at some stage the consideration
of the effects of matter fields. While there are many choices
for the matter fields and for their interactions, the simplest
actions to deal with in the framework of a lattice model for gravity
are the ones that represent one (or more) scalar fields [1].
In this talk I will briefly review these results.
Regge's lattice model for gravity is the natural discretization for
quantized gravity in four dimensions.
At the classical level, it is completely equivalent to general relativity,
and the correspondence is particularly transparent in the lattice
weak field expansion, with the invariant edge lengths playing
the role of infinitesimal geodesics in the continuum.
Recent work based on Regge's simplicial formulation of gravity has shown,
in pure gravity without matter,
the appearance in four dimensions of a phase transition in the
bare Newton's constant, separating a smooth phase with small negative
average curvature from a rough phase with large positive
curvature [2].
Here I will discuss a study we have performed to determine the nature
and size of the effects that appear when a scalar field is coupled
to gravity.
As will be discussed below, our results seem to indicate that the
'vacuum polarization' effects due to one single scalar field of small
mass are rather small for the observables we have investigated,
when compared to the dominant pure gravitational contribution.

\vskip 4.0mm
{\bf \noindent Action and Measure}
\vskip 2.0mm

For the gravitational field the following lattice action is used [3]
\beq
I_g [l] =  \sum_ {\rm hinges \, h } \Bigl [ \, \lambda \, V_h -
k \, A_h \d _h + a \, { A_ h^2  \delta _ h^2 \over  V _ h } \, \Bigr ] ,
\label{eq:acg}
\eeq
where $V_h$ is the volume per hinge (which is represented by a
triangle in four dimensions),
$A_h$ is the area of the hinge and $\delta_h$ the corresponding
deficit angle, proportional to the curvature at $h$.
All these quantities can be evaluated in terms of the lattice edge
lengths $l_{ij}$, which define the lattice geometry for a fixed
incidence matrix.
The underlying lattice is chosen to be hypercubic, with a natural simplicial
subdivision to make it rigid.
In the classical continuum limit the above action is then equivalent to
\beq
I_g [g] = \int  d^4 x  \, \sqrt g \, \Bigl [ \, \lambda - \half k \, R
+ \, \quarter a \, R _ { \mu \nu \rho \sigma }  R ^ { \mu \nu \rho \sigma }
+ \cdots \, \Bigr ] ,
\label{eq:acgc}
\eeq
with a cosmological constant term (proportional to $\lambda$), the
Einstein-Hilbert term ($k = 1 / 8 \pi G  $), and a higher derivative
term proportional to $a$.
For an appropriate choice of bare couplings,
the above lattice action is bounded below for a regular lattice, even
for $a=0$, due to the presence of a lattice momentum cutoff, which
cuts off the conformal mode fluctuations at high momenta.
For non-singular measures and in the presence of the $\lambda$-term
such a regular lattice can be shown to arise naturally.
The higher derivative terms can be set to zero ($a=0$),
but they nevertheless seem to be necessary for reaching the lattice continuum
limit, and are in any case generated by radiative corrections
already in weak coupling perturbation theory.

The scalar field $\phi_i$ is then defined at the vertices of the simplices.
One adds to the pure gravitational action the contribution
\beq
I_\phi [l, \phi ] = \half \sum_{<ij>} V_{ij} \,
\Bigl ( { \phi_i - \phi_j \over l_{ij} } \Bigr )^2 \, +
\half \sum_{i} V_{i} \, (m^2 + \xi R_i ) \phi_i^2  .
\label{eq:acp}
\eeq
The term containing the discrete analog of the scalar curvature involves
\beq
V_{i} R_i \equiv \sum_{ h \supset i } \delta_h A_h  \sim \sqrt{g} R ,
\eeq
and in the expression for the scalar action,
$V_{ij}$ is the volume associated with the edge $l_{ij}$,
while $V_i$ is associated with the site $i$.
Then the above scalar lattice action then corresponds to the
continuum expression
\beq
I_\phi [ g, \phi ] = \half \int \, \sqrt g \; [  \,
g ^ { \mu \nu } \, \partial _ \mu  \phi \, \partial _ \nu  \phi
+ ( m^2 + \xi  R ) \phi^2   ] + \cdots .
\eeq
The dimensionless coupling $\xi$ is arbitrary; in our work
we have only considered the case ($\xi = 0$) (minimal coupling).

The measure contains an integration over the scalar fields, and
an integration over the edge lengths. For the edge lengths we
write the lattice measure as [3]
\beq
\int d \mu _ \epsilon [ l ] =
\prod _ {\rm edges \, ij}  \int_ 0 ^ \infty \,
V_{ij}^{ 2 \sigma} \, { d  l _ {ij} ^ 2 } \, F [ l ] ,
\label{eq:meas}
\eeq
where $V_{ij}$ is the 'volume per edge',
$ F [l] $  is
a function of the edge lengths which enforces the higher-dimensional
analogs of the triangle inequalities,
and $\sigma = 0$ for the lattice analog of the DeWitt measure for pure
gravity.
In the presence of an $n_f$-component scalar field,
the power $\sigma$ needs to be changed,
and on the lattice one has $ \sigma  = n_f /30 $,
since with our discretization of spacetime based on the simplicial
subdivision of hypercubes
there are a total of
$2(2^d -1)=30$ edges emanating from each lattice vertex.
Note that {\it no} cutoff is imposed on small or large edge
lengths, if a non-singular measure such as $dl^2$ is used.
This fact is essential for the recovery of diffeomorphism invariance
close to the critical point, where on large lattices
a few rather long edges, as well as some rather short ones,
start to appear.

\vskip 4.0mm
{\bf \noindent Observables }
\vskip 2.0mm

When we consider gravity coupled to a scalar field, we can distinguish
two types of observables, those involving the metric field (the edge
lengths) only, and those involving also the scalar field.
The average curvature
\beq
\cR (\lambda, k, a) \, \sim \,
\, { < \int \sqrt{g} \, R > \over < \int \sqrt{g} > } ,
\eeq
belongs to the first class, and its lattice analog is defined as
\beq
{\cal R}(\lambda, k, a) \; = \;
<l^2> { < 2 \sum_h \delta_h A_h > \over < \sum_h V_h > } .
\label{eq:avr}
\eeq
In a similar way one can define the curvature fluctuation. It
is related to the (connected) scalar curvature
correlator at zero momentum
\beq
\chi_\cR \sim { \int d^4 x \int d^4 y < \sqrt{g} R (x) \sqrt{g} R (y) >_c
\over < \int d^4 x \sqrt{g} > } .
\eeq
A divergence in the curvature fluctuation is then indicative of
long range correlations (a massless graviton here).
Close to the critical point one expects for large separations a power
law decay in the geodesic distance,
\beq
< \sqrt{g} R (x) \sqrt{g} R (y) >
\mathrel{\mathop\sim_{ \vert x - y \vert \rightarrow \infty}}
\frac{1}{ \vert x-y \vert^{2n} } .
\eeq

In quantum gravity it is of interest to determine the
value of the low energy, renormalized coupling constants, and in
particular the effective cosmological constant $\lambda_{eff}$ and the
effective Newton's constant $ G_{eff} $.
Equivalently, one would like
to be able to determine the long distance limiting value of a
dimensionless ratio such as $\lambda_{eff} G_{eff}^2 $, and its
dependence on the linear size of the system $L = V^{1/4} $.
We have argued that the vacuum expectation value of the scalar
curvature can be used as
a definition of the effective, long distance cosmological constant
\beq
{\cal R} \, \sim \left ( { \lambda G } \right )_{eff} .
\label{eq:effr}
\eeq
Indeed in the pure gravity case there is evidence that the curvature
vanishes at a critical point in $G$, and for $G > G_c$ one finds
results which are consistent with a singularity of the type
\beq
\cR
\mathrel{\mathop\sim_{ G \rightarrow G_c}}
- A_\cR \, ( G - G_c )^\delta ,
\label{eq:rsing}
\eeq
where $\delta$ is a universal critical exponent [2].
If this is true, then $ (\lambda G )_{eff} \rightarrow 0 $ in
lattice units at the critical point.
While the location of the critical point $G_c$ and the amplitude
in general depend on the higher derivative coupling $a$ and other
non-universal parameters, the exponent is expected
to be universal. We have estimated its value at about 0.6.

One immediate consequence of this result is that in the smooth phase
with $ G > G_c$ the gravitational coupling
constant $G$ must increase with distance (anti-screening), at least
for rather short distances.
Introducing an arbitrary momentum scale $\mu $, one has close to the
ultraviolet fixed point the following short-distance behavior for
Newton's constant
\beq
G ( \mu ) - G_c \, = \, \left [  G ( \Lambda ) - G_c \right ]
\left ( \Lambda \over \mu \right ) ^{ 1 / \nu } ,
\label{eq:gscale}
\eeq
with $\Lambda $ the ultraviolet cutoff; the exponents $\delta$ and
$\nu$ are calculable and are related to each other via the scaling relation
$\nu = (1 + \delta ) / 4 \approx 0.4$.
Note that the opposite behavior (screening) would be true in the phase with
$G < G_c$, but such a phase is known not to be stable and leads to no
lattice continuum limit.

\vskip 4.0mm
{\bf \noindent Switching on the Scalar Field }
\vskip 2.0mm

In order to explore the ground state of four-dimensional simplicial
gravity coupled to matter beyond perturbation theory one has to resort
to numerical methods, where the edge lengths and scalars are updated by a
Monte Carlo method.
In our work we considered lattices with $4^4$ to $16^4$ sites.
As far as the properties of the critical point are concerned, one
still finds an apparently continuous phase transition between the smooth and
the rough phase, but with $A_{\cR} , G_c , \delta $ that will now
depend on $n_f$.
If one adopts the same procedure as for pure gravity, and fits the average
curvature for small scalar mass to an algebraic singularity, one finds
(for $a=0.005$ and $m=0.5$) $\delta=0.61(6)$, to be compared with the pure
gravity estimate $\delta=0.63(3)$ [2].
It appears therefore that, within errors, switching on the
scalar fields leaves the exponents almost unchanged.
For small non-integer $n_f$ we can write for the amplitude, critical
value of $k$ and the exponent in powers of the number of flavors $n_f$,
\bea
& A_{\cR} & =  A_0 + n_f A_1 + O(n_f^2) \nonumber \\
& G_c & =  G_0 + n_f G_1 + O(n_f^2) \nonumber \\
& \delta & =  \delta_0 + n_f \delta_1 + O(n_f^2) .
\eea
Our results seem to imply that the corrections due to the scalar
field are quite small, and that the coefficients
of the $n_f$ terms must be rather small. Since $G_c$ and $\delta$ are
almost unchanged, one can estimate $A_1$ in the following way.
One notes that for small $n_f$
the difference between the average curvature in the
presence of the
scalar field and in pure gravity determines the ratio of curvature
amplitudes $A_1/A_0$
\beq
{ \cR_{matter} \over \cR_{gravity} } \, = \,
{ \cR_{gravity+matter} - \cR_{gravity} \over \cR_{gravity} }
\mathrel{\mathop\sim_{ G \rightarrow G_c }}
{ A_1 \over A_0 } .
\eeq
Since the difference in the numerator is quite small, a very accurate
measurement of the average curvature in both cases is required. One
finds $ A_1 / A_0 \approx 0.053 / 3.79 = 0.014 $ for $a=0.005$.
A possible explanation for the smallness of this ratio
can be found in the Nielsen-Hughes formula, where the strength
of the relative contributions (here gravity versus a scalar)
is determined by the particle's relative spin.

Since the effects of the scalar fields are quite small, one
can discuss the renormalization properties of the gravitational
coupling constants without distinguishing the two cases explicitly.
Let us address here briefly the renormalization properties of
the couplings $ G$ and $\lambda$. Consider a universe of finite
linear extent $L$, and set the graviton mass equal to the inverse
of this size (since essentially the associated correlation length
$\xi$ saturates at the system size, $ \xi \sim (G - G_c)^{-\nu} \sim L $).

{}From Eq. ~(\ref{eq:gscale}) one has then the following size dependence
for the dimensionful Newton's constant, valid for `short' distances
$1 / \mu \ll L $,
\beq
G_{eff} ( \mu ) \, \mathrel{\mathop\sim_{ L , \; 1 / \mu \gg l_0 }} \,
l_0^{2} G_c + l_0^{2} \left ( { 1 \over \mu L } \right )^{ 1 / \nu}
\eeq
(with $ 1 / \nu \approx 2.46 $),
while from Eqs. ~(\ref{eq:effr}) and ~(\ref{eq:rsing}) one obtains
\beq
\lambda_{eff} ( \mu ) \, \mathrel{\mathop\sim_{ L , \; 1 / \mu \gg l_0 }} \,
l_0^{-4} \left ( { \mu l_0 } \right )^{ 4 - 1 / \nu}
\left [ G_c + \left ( { 1 \over \mu L } \right )^{ 1 / \nu} \right ]^{-1} .
\eeq
Here again $l_0$ is of the order of the average lattice spacing, and we have
restored the correct dimensions for $G_{eff}$ and $ \lambda_{eff} $.
For the dimensionless ratio $ (G^2 \lambda)_{eff}$ one then finds that it
can be made very small, provided the linear size $L$ is large enough.

\vskip 4.0mm
{\bf \noindent Conclusions }
\vskip 2.0mm

Our work up to now suggests that the feedback of the scalar fields on
the geometry is quite small on purely gravitational quantities,
such as the average curvature.
It appears therefore that the approximation in which matter internal
loops are neglected (quenched approximation) could be considered a
reasonable starting point, and that quantities such as the critical exponents
should not be too far off in this case.
To the extent that the coupling between the scalar and metric
degrees of freedom is weak close to the critical point, we have argued
that gravity is indeed weak.
Our results for the exponents and the overall phase diagram seem to suggest
that the gravitational coupling exhibits an infrared growth
away from the fixed point, of the
type $ G ( \mu ) \sim \mu^{- 1 / \nu }$, while for the cosmological
constant we have found a decrease in the infrared,
$ \Lambda (\mu ) \sim \mu^{ 4 - 1 / \nu }$, with an exponent
$\nu$ given approximately by $\nu \approx 0.4$, and only weakly
dependent on the matter content.

\vskip 4.0mm
{\bf \noindent Acknowledgements}
\vskip 2.0mm

The work presented here was done in collaboration with Ruth M. Williams.
Numerical computations were performed at NCSA under
a {\sl Grand Challenge} allocation grant.
This work was supported in part by the NSF under grant PHY-9208386.
During the conference the author has benefited from conversations
with H. Nielsen, P. Menotti and J. Smit.
\vskip 4.0mm
{\bf \noindent References}
\vskip 2.0mm
\noindent
[1] H.W. Hamber and R.M. Williams, preprint DAMTP-93-27/UCI-93-16.
\newline
\noindent
[2] H.W. Hamber, {\it Phys.\ Rev.}  {\bf D45} (1992) 507; and
{\it Nucl.\ Phys.} {\bf B400} (1993) 347.
\newline
\noindent
[3] H.W. Hamber and R.M. Williams, {\it Nucl.\ Phys.} {\bf B248} (1984)
392; {\bf B260} (1985) 747;  {\bf B269} (1986) 712;
H.W. Hamber, in {\sl Les Houches 1984}, Session XLIII,
(North Holland, 1986), pp. 375-439.

\end{document}